\documentclass[letterpaper,twocolumn,pre,aps,showpacs,floatfix,superscriptaddress,nofootinbib]{revtex4-1}
\usepackage[latin1]{inputenc}
\usepackage{bm}
\usepackage{multirow}
\usepackage{amssymb}
\usepackage{amsbsy}
\usepackage{amsmath}
\usepackage{graphicx}
\usepackage{epsfig}
\usepackage{placeins}
\usepackage{ulem}
\usepackage[usenames,dvipsnames]{color}
\usepackage[colorlinks,linkcolor=blue,citecolor=blue,urlcolor=blue]{hyperref}
\makeatletter

\begin{document} 
\title{Real-time broadening of non-equilibrium density profiles and\\
the role of the specific initial-state realization}

\author{R. Steinigeweg}
\email{rsteinig@uos.de}
\affiliation{Department of Physics, University of Osnabr\"uck, D-49069 Osnabr\"uck, Germany}

\author{F. Jin}
\affiliation{Institute for Advanced Simulation, J\"ulich Supercomputing Centre, Forschungszentrum J\"ulich, D-52425 J\"ulich, Germany}

\author{D. Schmidtke}
\affiliation{Department of Physics, University of Osnabr\"uck, D-49069 Osnabr\"uck, Germany}

\author{H. De Raedt}
\affiliation{Zernike Institute for Advanced Materials, University of Groningen, NL-9747AG Groningen, The Netherlands}

\author{K. Michielsen}
\affiliation{Institute for Advanced Simulation, J\"ulich Supercomputing Centre, Forschungszentrum J\"ulich, D-52425 J\"ulich, Germany}
\affiliation{RWTH Aachen University, D-52056 Aachen, Germany}

\author{J. Gemmer}
\email{jgemmer@uos.de}
\affiliation{Department of Physics, University of Osnabr\"uck, D-49069 Osnabr\"uck, Germany}

\begin{abstract}
The real-time broadening of density profiles starting from non-equilibrium
states is at the center of transport in condensed-matter systems and dynamics
in ultracold atomic gases. Initial profiles close to equilibrium are expected
to evolve according to linear response, e.g., as given by the current
correlator evaluated exactly at equilibrium. Significantly off equilibrium,
linear response is expected to break down and even a description in terms
of canonical ensembles is questionable. We unveil that single pure states
with density profiles of maximum amplitude yield a broadening in perfect
agreement with linear response, if the structure of these states involves
randomness in terms of decoherent off-diagonal density-matrix elements. While
these states allow for spin diffusion in the XXZ spin-$1/2$ chain at large
exchange anisotropies, coherences yield entirely different behavior.
\end{abstract}


\maketitle

\section{Introduction}

The mere existence of equilibration and thermalization
is a key issue in many areas of modern many-body physics. While this question
has a long and fertile history, it has experienced an upsurge of interest in
recent years \cite{eisert2015} due to the advent of cold atomic gases
\cite{langen2015} as well as due to the discovery of new states of matter such
as many-body localized phases \cite{nandkishore2015}. In particular, the
theoretical understanding has seen substantial progress by the fascinating
concepts of eigenstate thermalization \cite{deutsch1991, srednicki1994,
rigol2008} and typicality of pure quantum states \cite{gemmer2003,
goldstein2006, popescu2006, reimann2007, bartsch2009, sugiura2012, sugiura2013,
elsayed2013} as well as by the invention of powerful numerical methods such as
density-matrix renormalization group \cite{schollwoeck2005}. Much less is known
on the route to equilibrium as such \cite{reimann2016} and still the derivation
of the conventional laws of (exponential) relaxation and (diffusive) transport
on the basis of truly microscopic principles is a challenge to theory
\cite{buchanan2005}.

In strictly isolated systems any coupling to heat baths or particle reservoirs
and any driving by external forces is absent. In such systems, the only
possibility to induce a non-equilibrium process is the preparation of a proper
initial state. While different ways of preparation can be chosen, a sudden
quench of the Hamiltonian is a common preparation scheme  \cite{essler2016}.
However, once a specific state is selected, a crucial question is: To
what extent is this state a non-equilibrium state? To answer this question, it
is natural to measure the observable one is interested in. If the expectation
value is far from equilibrium, the state should be also. If this value is close
to equilibrium, the state should be correspondingly. Moreover, only in the
latter case, the resulting dynamics of the expectation value and linear response
theory are expected to agree with each other. While this line of reasoning is
certainly intuitive, it neglects internal degrees of freedom of the initial
state. In particular, the measurement of a single observable cannot detect
if the underlying state is pure or mixed, entangled or non-entangled, etc.
Therefore, an intriguing question is: Do such internal details play any
role for the dynamics of an expectation value?

In this paper, we investigate exactly this question for the anisotropic
spin-$1/2$ Heisenberg chain. Dynamics in this integrable many-body model
has been under active scrutiny in various theoretical works and, in
particular, spin dynamics constitutes a demanding problem resolved only
partially despite much effort \cite{shastry1990, narozhny1998, zotos1999,
benz2005, heidrichmeisner2003, fujimoto2003, prosen2011, prosen2013,
herbrych2011, karrasch2012, karrasch2013, steinigeweg2014-1, steinigeweg2015,
carmelo2015, gobert2005, sirker2009, grossjohann2010, prelovsek2004,
znidaric2011, steinigeweg2011, karrasch2014, steinigeweg2012}, even within the
linear response regime and at high temperatures. While it has become clear
that quasi-local conservation laws \cite{prosen2011, prosen2013}
necessarily lead to ballistic behavior below the isotropic point,
numerical studies \cite{prelovsek2004, znidaric2011, karrasch2014,
steinigeweg2011} have reported signatures of diffusion above this point, in
agreement with perturbation theory \cite{steinigeweg2011} and classical
simulations \cite{steinigeweg2012}.

To investigate spin transport, we first introduce a class of pure initial
states. These initial states feature identical density profiles, where a
maximum $\delta$ peak is located in the middle of the chain and lies on
top of a homogeneous background, similar to \cite{karrasch2014}. For a subclass
with internal randomness we then show analytically that the resulting
non-equilibrium dynamics can be related to equilibrium correlation functions
via the concept of typicality. This relation is verified in addition by
large-scale numerical simulations. These numerical simulations also unveil the
existence of remarkably clean diffusion for large exchange anisotropies, as one
of our central findings. Eventually, we demonstrate that entirely different
behavior emerges without any randomness in the initial state.

\section{Model and Observables}

The Hamiltonian of the XXZ spin-$1/2$ chain with periodic boundary conditions
reads
\begin{equation}
H = J \sum \limits^L_{r=1} (S^x_r S^x_{r+1} + S^y_r S^y_{r+1} + \Delta S^z_r
S^z_{r+1}) \, ,
\end{equation}
where $S_r^{x,y,z}$ are spin-$1/2$ operators at site $r$, $L$ is the number of
sites, $J > 0$ is the antiferromagnetic exchange coupling constant, and $(\Delta
-1)$ is the anisotropy. For all parameters, this model is integrable in terms of
the Bethe Ansatz and the total magnetization $S^z = \sum_r S_r^z$ is a strictly
conserved quantity. We take into account all subsectors of $S^z$, i.e.,
we consider the case $\langle S^z \rangle = 0$. We note that, via the
Jordan-Wigner transformation, this model can be mapped onto a chain of spinless
fermions with particle interactions of strength $\Delta$ and total particle
number $N = S^z + L/2$, i.e., $\langle N \rangle = L/2$ (see Appendix \ref{hf}
for the half-filling case $N = L/2$).

We are interested in the non-equilibrium dynamics of the local occupation
numbers $n_r = S_r^z + 1/2$. Specifically, we consider the expectation values
$p_r(t) = \text{tr} [ n_r \, \rho(t) ]$ for the density matrix $\rho(t)$ at time
$t$. In this way, we study the time-dependent broadening of density profiles 
for a given initial state $\rho(0)$. In this paper, we focus on pure states
$\rho(0) = | \psi(0) \rangle \langle \psi(0) |$.

\section{Initial States}

Obviously, it is possible to choose many different initial
states $|\psi (0) \rangle$ and the resulting dynamics can depend on details of
the specific choice. A frequently used preparation scheme is a quantum quench,
i.e., $|\psi (0) \rangle$ is the eigenstate of another Hamiltonian. In this
paper, however, we proceed in a different way.

To introduce our class of initial states, let $| \varphi_k \rangle$ be the
common eigenbasis of all $n_r$, i.e., the Ising basis. Then, this class reads
\begin{equation}
| \psi(0) \rangle \propto n_{L/2} \, | \Phi \rangle \, , \quad | \Phi \rangle =
\sum_{k=1}^{2^L} c_k \, | \varphi_k \rangle \, ,
\end{equation}
where $c_k$ are complex coefficients and $n_{L/2}$ projects onto Ising states
with a particle in the middle of the chain. By construction, $p_{L/2}(0) = 1$
is maximum.

In the above class, a particular state is the one where all $c_k$ are the same.
It yields $p_{r \neq L/2}(0) = p_\text{eq.} = 1/2$ and still $p_{L/2}(0) = 1$.
Hence, its density profile has a $\delta$ peak on top of a homogeneous
background. However, exactly this density profile also results when the $c_k$
are drawn at random according to the unitary invariant Haar measure
\cite{bartsch2009} (where real and imaginary part of the $c_k$ are drawn from
a Gaussian distribution with zero mean, as done in our numerical simulations
perfomed below). In other words, it is impossible to distinguish the two states
with equal and random coefficients by a measurement of their initial density
profiles $p_r(0)$ \cite{note}. Only at times $t > 0$, their density profiles
$p_r(t)$ can be different, if these density profiles differ at all. Note that
similar $p_r(0)$ have been studied in Ref.\ \cite{karrasch2014}.

Because our initial states are pure and have maximum $p_{L/2}(0) = 1$ as well,
these states have to be considered as far-from-equilibrium states. Thus, it is
natural to expect that the resulting dynamics of $p_r(t)$ cannot be described by
linear response theory. However, such a expectation turns out to be wrong for
the case of random $c_k$. In this case, $| \Phi \rangle$ is a typical
state \cite{gemmer2003,
goldstein2006, popescu2006, reimann2007, bartsch2009, sugiura2012, sugiura2013,
elsayed2013}, i.e., a trace $\text{tr}[ \bullet ]$ can be approximated by the
expectation value $\langle \Phi | \bullet | \Phi \rangle$ with high accuracy in
large Hilbert spaces. Using this fact and exact math (see Appendix \ref{ta} for
more details), we find the relation
\begin{equation}
p_r(t) -  p_\text{eq.} = 2 \, \langle (n_{L/2} - p_\text{eq.})(n_r(t) -
p_\text{eq.}) \rangle \, , \label{typicality}
\end{equation}
where $\langle \bullet \rangle = \text{tr}[\bullet] / 2^L$. This relation is a
first main result of our paper. It unveils that the expectation value $p_r(t)$
of a far-from-equilibrium state is directly connected to a equilibrium
correlation function. It is important to note that such a relation cannot be
derived for the other case of equal $c_k$ (see also Appendix \ref{stor} for
the specific type of randomness).

Due to the above relation, it is also possible to connect our non-equilibrium
dynamics to the Kubo formula. To this end, one has to define the spatial
variance
\begin{equation}
\sigma(t)^2 = \sum_{r=1}^L r^2 \, \delta p_r(t) - \Big[ \sum_{r=1}^L r \, \delta
p_r(t) \Big ]^2 \label{variance}
\end{equation}
with $\delta p_r(t) = 2(p_r(t) - p_\text{eq.})$ and $\sum_{r=1}^L \delta p_r(t)
= 1$. Then, following Ref.\ \cite{steinigeweg2009}, it is straightforward to
show that the time derivative of this variance
\begin{equation}
\frac{\text{d}}{\text{d}t} \sigma(t)^2 = 2 \, D(t) \label{derivative}
\end{equation}
is given by the time-dependent diffusion coefficient
\begin{equation}
D(t) = \frac{4}{L} \int_0^t \text{d}t' \, \langle j(t') j \rangle \, , \label{D}
\end{equation}
where $j = \sum_{r=1}^L S_r^x S_{r+1}^y - S_{r}^y S_{r+1}^x$ is the well-known
spin current. For $\Delta = 0$, $[j,H] = 0$ leads to ${\cal D}(t) \propto t$
such that $\sigma(t) \propto t^2$ scales ballistically. The partial
conservation of $j$ for $\Delta < 1$ \cite{shastry1990, narozhny1998, zotos1999,
benz2005, heidrichmeisner2003, fujimoto2003, prosen2011, prosen2013, herbrych2011,
karrasch2012, karrasch2013, steinigeweg2014-1, steinigeweg2015} also excludes
diffusive scaling $\sigma(t)^2 \propto t$ in this $\Delta$ regime. In fact,
signatures of diffusion at high temperatures have been found only in the regime
of large anisotropies $\Delta > 1$ \cite{prelovsek2004, znidaric2011, steinigeweg2011,
karrasch2014}. Note that $\sigma(t)^2 \propto t$ is merely a necessary and no
sufficient criterion for diffusion since, by definition, the variance yields no
information beyond the width of the distribution $\delta p_r(t)$. This is why we
study the full space dependence. For a recent numerical survey of Eq.\
(\ref{derivative}), see \cite{karrasch2016}.

\begin{figure}[t]
\includegraphics[width=0.95\columnwidth]{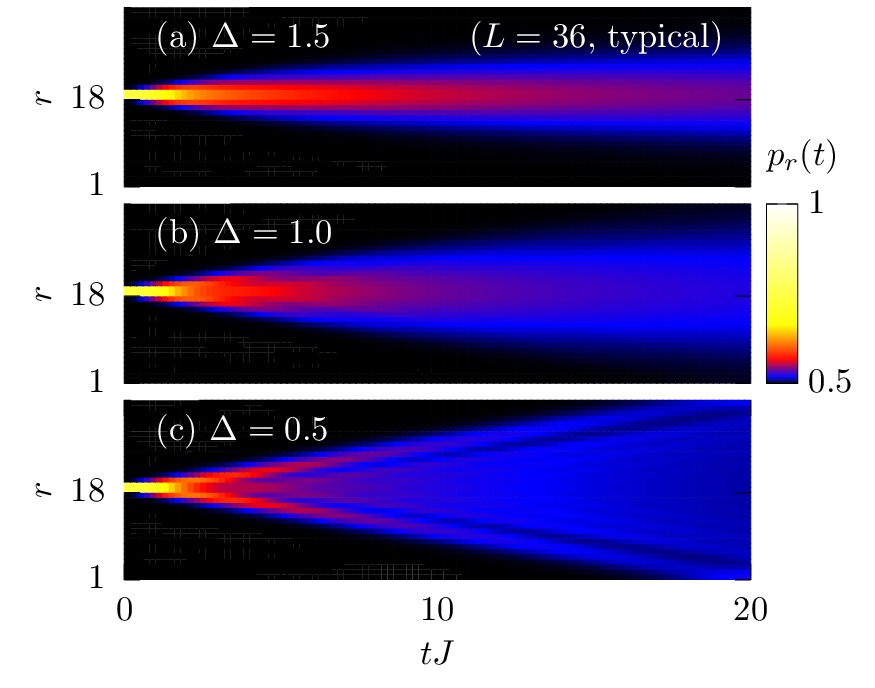}
\caption{(Color online) Time-space density plot of occupation numbers $p_r(t)$
for a {\it typical} initial state $|\psi(0) \rangle$ in the XXZ spin-1/2 chain
with $L=36$ sites and different anisotropies: (a) $\Delta = 1.5$, (b) $\Delta
= 1.0$, (c) $\Delta = 0.5$. The broadening in (a) is consistent with a diffusive
process while the broadening in (c) is ballistic.
}
\label{Fig1}
\end{figure}

\section{Numerical Method and Results}

Numerically, the time evolution of a pure
state $| \psi(t) \rangle$ can be calculated by the method of full exact
diagonalization. But this method is restricted to $L \sim 20$ sites, even
if symmetries such as the translation invariance of $H$ are taken into account.
Thus, we proceed differently and rely on a forward propagation of $|\psi (t)
\rangle$ in real time. Such a propagation can be done by the use of
fourth-order Runge-Kutta \cite{steinigeweg2014-1, steinigeweg2015, elsayed2013}
or more sophisticated schemes such as Trotter decompositions or Chebyshev
polynomials \cite{steinigeweg2014-2, jin2015}. Here, we use a massively
parallelized implementation of a Chebyshev-polynomial algorithm. In this way,
we can treat system sizes as large as $L=36$. For such $L$, we can guarantee
that the initial $\delta$ peak is located sufficiently far from the boundary
of the chain. Otherwise, we would have to deal with trivial finite-size effects
and also Eq.\ (\ref{derivative}) would not hold \cite{steinigeweg2009}.

Next, we turn to our numerical results, starting with a typical initial state
$|\psi (0) \rangle$, i.e., the case of random $c_k$. For a single realization
of this state, we summarize in Fig.\ \ref{Fig1} the resulting expectation value
$p_r(t)$ in a 2D time-space density plot for different anisotropies $\Delta =
1.5$, $1.0$, $0.5$ and a large system with $L = 36$ sites. Several comments are
in order. First, for all values of $\Delta$ shown, the initial $\delta$ peak
monotonously broadens as a function of time and the non-equilibrium density
profiles have the irreversible tendency to equilibrate. Such equilibration is
non-trivial in view of our isolated and integrable model. Second, for times
below the maximum $t J = 20$ depicted, the spatial extension of the density
profiles is still smaller than the length of the chain. Thus, unwanted boundary
effects do not emerge for such times. Third, the broadening of the density
profiles is faster for smaller values of $\Delta$ because the scattering due to
particle interactions decreases as $\Delta$ decreases. Moreover, for the small
$\Delta = 0.5$ in Fig.\ \ref{Fig1} (c), the width of the density profile clearly
increases linearly as a function of time. This linear increase is the expected
ballistic dynamics arising from partial conservation of the spin current. In
contrast, for the larger $\Delta = 1.5$ and $1.0$ in Figs.\ \ref{Fig1} (a) and
(b), the width of the density profiles does not increase linearly and is
rather reminiscent of a square-root behavior. However, such a conclusion is
not possible on the basis of a density plot.

\begin{figure}[t]
\includegraphics[width=0.95\columnwidth]{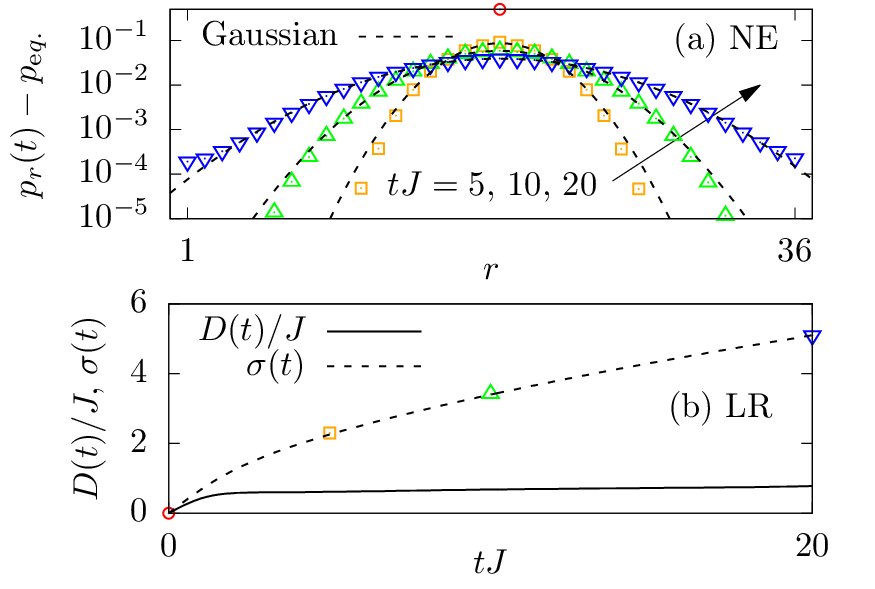}
\caption{(Color online) (a) Density profile $p_r(t)$ with respect to site $r$
at fixed times $t J = 0$, $5$, $10$, $20$ for a single anisotropy $\Delta =
1.5$ (and the parameters in Fig.\ \ref{Fig1}), shown in a semi-log plot
(symbols). The Gaussian fits indicated describe the data very well over several
orders of magnitude (curves). (b) Time dependence of diffusion coefficient
$D(t)$ and profile width $\sigma(t)$ according to linear response theory,
calculated in Ref.\ \cite{steinigeweg2015} for the same anisotropy $\Delta =
1.5$ and $L=34$ sites (curves). For comparison, the standard deviation
$\sigma(t)$ of the Gaussian fits in (a) is depicted (symbols).}
\label{Fig2}
\end{figure}

To gain insight into the dynamics at $\Delta = 1.5$, we depict in Fig.\
\ref{Fig2} (a) the site dependence of the expectation values $p_r(t)$ at fixed
times $t J = 0$, $5$, $10$ and $20$. Conveniently, we subtract the equilibrium
value $p_\text{eq.}$ and use a semi-log plot to visualize also the tails of the
density profiles. As illustrated by fits, the site dependence can be described
by Gaussians (with $\sigma_\text{f}(t)$ as the only fit parameter)
\begin{equation}
p_r(t) - p_\text{eq.} = \frac{1}{2} \, \frac{1}{\sqrt{2 \pi} \, \sigma_\text{f}(t)}
\, \exp \! \left[-\frac{(r - L/2)^2}{2 \, \sigma_\text{f}(t)^2} \right ]
\end{equation}
and, remarkably, over several orders of magnitude. Such a pronounced Gaussian
form of the density profiles is a second main result of our paper and has, to
best of our knowledge, not been reported in the literature yet. This result
unveils that the standard deviation $\sigma_\text{f}(t)$ is not just a width but
also the only parameter required to describe the full site dependence. Furthermore,
the Gaussian form is one of the clearest signatures of diffusion so far. Still,
diffusion requires that $\sigma_\text{f}(t)$ scales as $\sigma_\text{f}(t) \propto
\sqrt{t}$.

To further judge on diffusion, we show in Fig.\ \ref{Fig2} (b) the standard
deviation $\sigma_\text{f}(t)$, as resulting from the Gaussian fits in Fig.\ \ref{Fig2}
(a). We further depict linear-response results for $\sigma(t)$ in Eq.\
(\ref{derivative}) and the underlying $D(t)$ in Eq.\ (\ref{D}), as calculated in
Ref.\ \cite{steinigeweg2015} for $L=34 \sim 36$. On the one hand, the excellent
agreement shows the very high accuracy of the typicality relation in Eq.\
(\ref{typicality}). On the other hand, this agreement demonstrates that the
known linear-response result $\sigma(t) \propto \sqrt{t}$, resulting from $D(t)
\approx \text{const.}$ at such $t$ \cite{steinigeweg2011, steinigeweg2015,
karrasch2014}, also holds for our non-equilibrium density dynamics. Hence,
together with the Gaussian form, we can conclude that diffusion exists.

\begin{figure}[t]
\includegraphics[width=0.95\columnwidth]{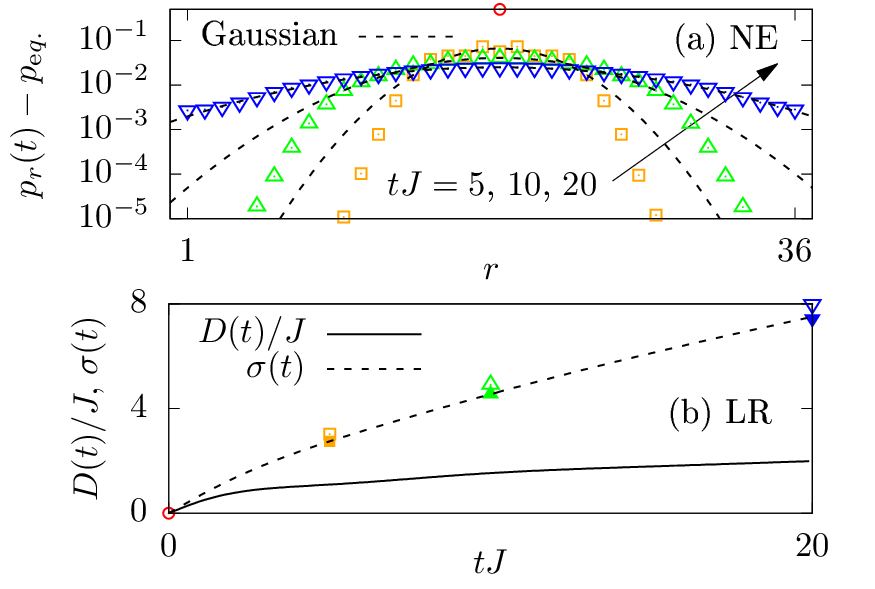}
\caption{(Color online) The same data as depicted in Fig.\ \ref{Fig1} but now
for the anisotropy $\Delta = 1.0$. In (a) the Gaussian fits cannot describe the
tails of the density profiles accurately. In (b) the standard deviation of these
fits (open symbols) and according to Eq.\ (\ref{variance}) (closed symbols)
still agrees with linear response; however, the time dependence is clearly
inconsistent with diffusion. Note that finite-size effects are negligibly
small, see Appendix \ref{fse}.}
\label{Fig3}
\end{figure}

An analogous analysis for the isotropic point $\Delta = 1.0$ in Fig.\ \ref{Fig3}
(a) shows that simple Gaussians are not able to describe the tails of the
density profiles accurately. This is why the standard deviation $\sigma_\text{f}(t)$
of corresponding fits slightly deviates from the linear-response result in Fig.\
\ref{Fig3} (b). But these deviations disappear if $\sigma(t)$ is calculated
exactly according to Eq.\ (\ref{variance}).  Most notably, however, the time
dependence of $\sigma(t)$ is inconsistent with diffusion, as can be seen
easiest from the non-constant $D(t)$. In fact, $\sigma(t)$ points to
superdiffusion \cite{znidaric2011, steinigeweg2012}, contrary to
\cite{khait2016}.

\begin{figure}[t]
\includegraphics[width=0.95\columnwidth]{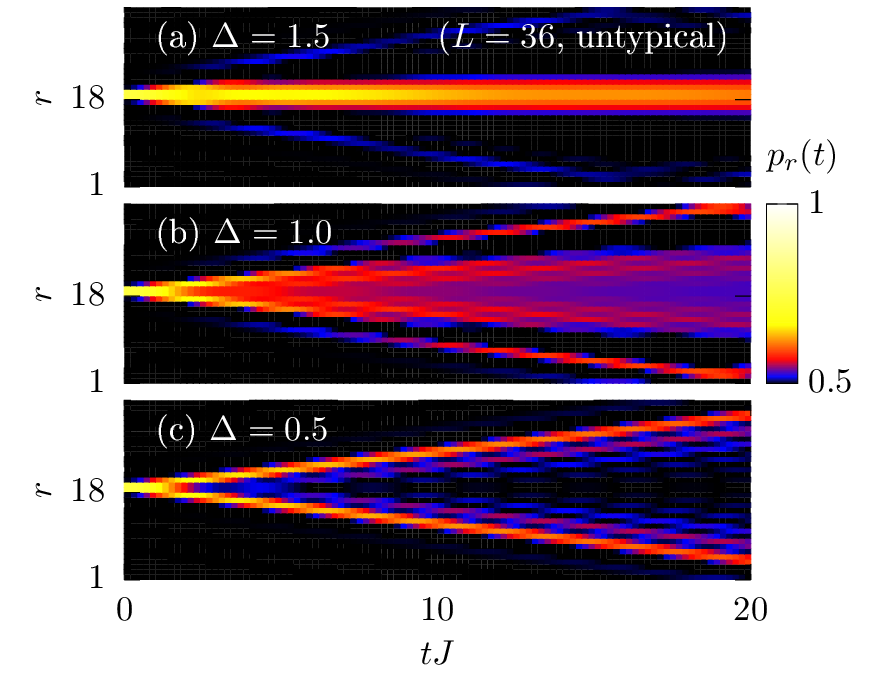}
\caption{(Color online) Time-space density plot of occupation numbers $p_r(t)$
for another and {\it untypical} initial state $|\psi(0) \rangle$ in the XXZ
spin-1/2 chain with $L=36$ sites and different anisotropies: $\Delta = 1.5$, (b)
$\Delta = 1.0$, (c) $\Delta = 0.5$. Compared to Fig.\ \ref{Fig1}, the dynamics
is frozen in (a), similar to \cite{gobert2005}, and features pronounced jets
in (c).}
\label{Fig4}
\end{figure}

Now, we turn to the untypical initial state $| \psi(0) \rangle$, i.e., the case
of equal $c_k$. Recall that for this state we obtain the same initial density
profile but the relations in Eqs.\ (\ref{typicality}) and (\ref{derivative}) do
not need to hold. In Fig.\ \ref{Fig4} we summarize the resulting expectation
values $p_r(t)$ in a 2D time-space density plot again. Compared to Fig.\
\ref{Fig1}, the broadening turns out to be clearly different. The dynamics is
frozen for $\Delta = 1.5$ in Fig.\ \ref{Fig4} (a)  and features pronounced jets
for $\Delta = 0.5$ in Fig.\ \ref{Fig4} (c). In particular, we do not find obvious
indications of equilibration, at least for all times considered. These
observations constitute a third main result of our paper. This result
suggests that the lack of internal randomness in the initial condition is
essential for the observation of non-equilibrium dynamics beyond linear response
theory.

Finally, let us briefly mention another property of the untypical initial state
$| \psi(0) \rangle$, which could be responsible for the special dynamics found.
This property is the lack of entanglement. In fact, it is easy to see that
$|\psi(0) \rangle$ can be written as the product state
\begin{equation}
| \psi(0) \rangle \propto \ldots (| \! \uparrow \rangle + | \! \downarrow \rangle)
\, \otimes \, | \! \uparrow \rangle \, \otimes \, (| \! \uparrow \rangle + | \!
\downarrow \rangle) \ldots \label{productstate}
\end{equation}
with a spin-up state $| \! \uparrow \rangle$ in the middle of the chain and a
spin-up/spin-down superposition $| \! \uparrow \rangle + | \! \downarrow
\rangle$ at all other sites. By definition, such a product state is not
entangled at all. In clear contrast, the typical initial state cannot be written
as a product state.

\section{Conclusions}

In this paper, we have investigated the real-time broadening
of non-equilibrium density profiles in the spin-$1/2$ XXZ chain. First, we have
introduced a class of pure initial states with identical density profiles
where a maximum $\delta$ peak is located in the middle of the chain. Then, we
have shown for a subclass with internal randomness that the resulting
non-equilibrium dynamics can be connected to equilibrium correlation functions
via the concept of typicality. This analytical result has been also verified by
large-scale numerical simulations. These numerical simulations have further
unveiled the existence of diffusion for large exchange anisotropies, as one of
our key results. Finally, we have demonstrated that entirely different behavior
emerges without any randomness in the initial state. Promising future directions
of research include the identification of typical and untypical initial states in
non-integrable models, in many-body localized phases, and at low temperatures
as well as a systematic analysis of the role of entanglement.

\section*{Acknowledgments}

We sincerely thank T.\  Prosen and F.\ Heidrich-Meisner
for fruitful discussions. In addition, we gratefully acknowledge the computing
time, granted by the ``JARA-HPC Vergabegremium'' and provided on the ``JARA-HPC
Partition'' part of the supercomputer ``JUQUEEN'' \cite{stephan2015} at
Forschungszentrum J\"ulich.

\appendix

\section{Half-Filling Sector} \label{hf}

\begin{figure}[t]
\includegraphics[width=0.95\columnwidth]{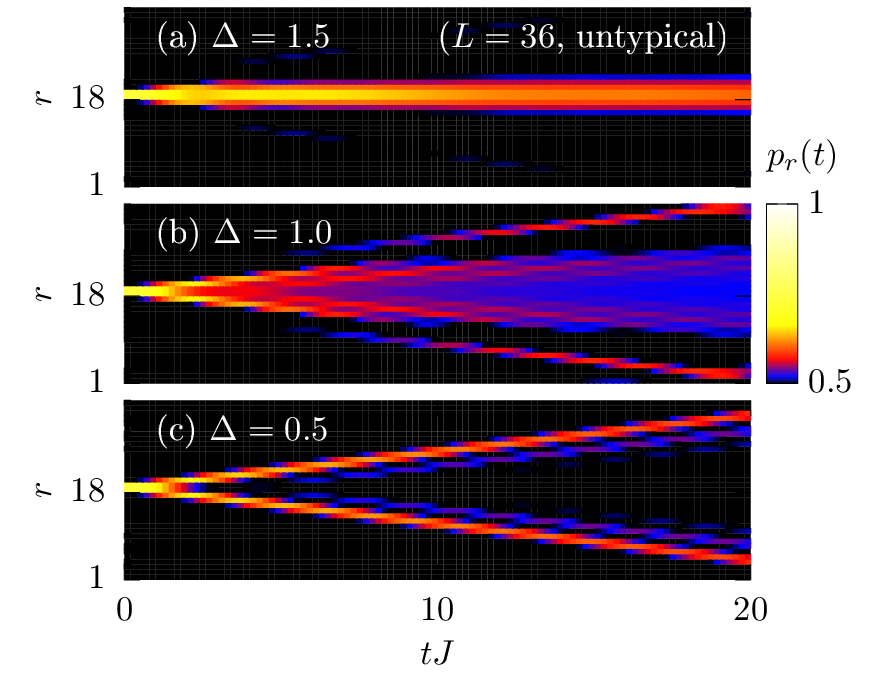}
\caption{(Color online) The same data as shown in Fig.\ 4
but now for the half-filling sector $S^z = 0$.}
\label{Fig5}
\end{figure}

To demonstrate that our results do not depend on our specific choice of $\langle
S^z \rangle = 0$, we do the calculation in, e.g., Fig.\ 4 again for the
half-filling sector $S^z = 0$. We depict the corresponding results in Fig.\
\ref{Fig5}. It is clearly visible that the real-time broadening of the
expectation values $p_r(t)$ is practically the same, apart from minor details
related to $p_\text{eq.} \approx 1/2$ in the half-filling case.

\section{Typicality Approximation} \label{ta}

Here, we provide details on the calculation leading to the relation in Eq.\
(3) of the main text. By carrying out the multiplication of the
two brackets in the correlation function
\begin{equation}
C(t) = 2 \, \langle (n_{L/2} - p_\text{eq.})(n_r(t) - p_\text{eq.}) \rangle
+ p_\text{eq.} \label{T1}
\end{equation}
and applying $\langle n_r(t) \rangle = p_\text{eq.}$, we obtain
\begin{equation}
C(t) = 2 \, \langle n_{L/2} \, n_r(t) \rangle = 2 \, \frac{\text{tr}[n_{L/2}
\, n_r(t)]}{2^L} \, .
\end{equation}
Using $n_{L/2}^2 = n_{L/2}$ and a cyclic permutation
in the trace, we get
\begin{equation}
C(t) = 2 \, \frac{\text{tr}[n_{L/2} \, n_r(t) \, n_{L/2}]}{2^L}
\, .
\end{equation}
Exploiting typicality of the pure state $| \Phi \rangle$, the correlation
function can be rewritten as
\begin{equation}
C(t) = 2 \, \frac{\langle \Phi | \, n_{L/2} \, n_r(t) \, n_{L/2} \, | \Phi
\rangle}{\langle \Phi | \Phi \rangle} + \epsilon
\end{equation}
with the small error $\epsilon \propto 2^{-L/2}$. Due to $n_{L/2}^\dagger
= n_{L/2}$, this expression becomes
\begin{equation}
C(t) = 2 \, \frac{\langle n_{L/2} \, \Phi | \, n_r(t) \, | n_{L/2} \, \Phi
\rangle} {\langle \Phi | \Phi \rangle} \, + \epsilon
\end{equation}
and, due to $n_r(t) = e^{\imath H t} \, n_r \, e^{-\imath H t}$, it reads
\begin{equation}
C(t) = \frac{\langle e^{-\imath H t} \, n_{L/2} \, \Phi | \, n_r \, |
e^{-\imath H t} \, n_{L/2} \, \Phi \rangle} {\langle \Phi | \Phi \rangle/2} \,
+ \epsilon \, ,
\end{equation}
where we have moved in addition the factor $2$ from the front to the
denominator. Finally, due to the definition of $| \psi(0) \rangle$, we can
write
\begin{equation}
C(t) = \langle \psi(t) | \, n_r \, | \psi(t) \rangle + \epsilon = p_r(t) +
\epsilon \, . \label{T2}
\end{equation}
Therefore, comparing Eqs.\ (\ref{T1}) and (\ref{T2}) and skipping the small
error $\epsilon$ for clarity yields
\begin{equation}
p_r(t) -  p_\text{eq.} = 2 \, \langle (n_{L/2} - p_\text{eq.})(n_r(t) -
p_\text{eq.}) \rangle \, .
\end{equation}

\begin{figure}[t]
\includegraphics[width=0.95\columnwidth]{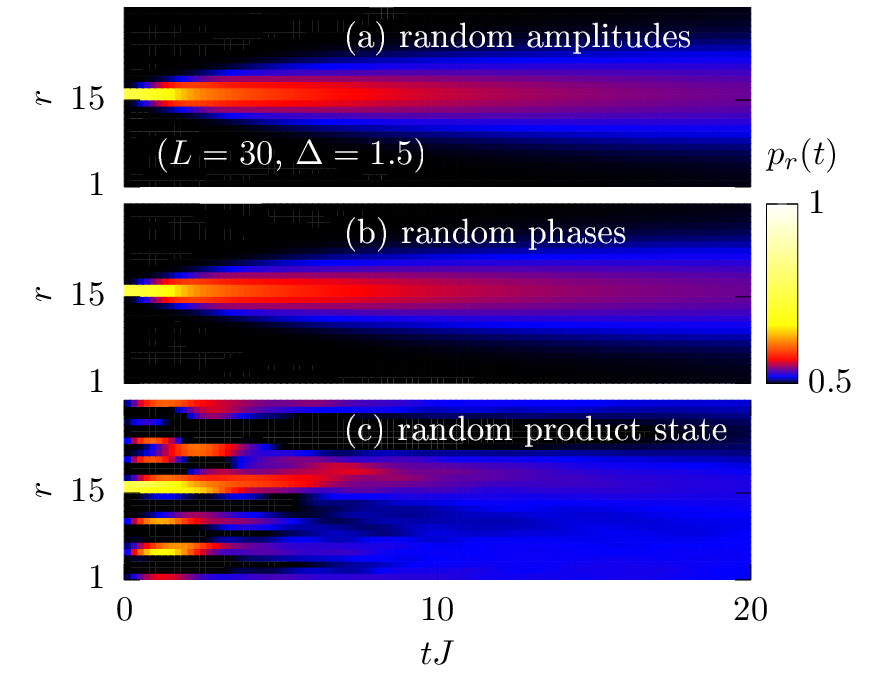}
\caption{(Color online) Time-space density plot of occupation numbers $p_r(t)$
in the spin-$1/2$ XXZ chain with $L=30$ sites and a single anisotropy $\Delta =
1.5$ for three different types of randomness in the pure initial state: (a)
random amplitudes, (b) random phases, (c) random product state; see text for the
detailed definitions.}
\label{Fig6}
\end{figure}

\section{Specific Type of Randomness} \label{stor}

As stated in the main text, the relations in Eqs.\ (3) and
(5) have to be understood for typical states $| \Phi \rangle$
drawn at random according to the unitary invariant Haar measure (where real
and imaginary part of the $c_k$ are drawn from a Gaussian distribution with
zero mean). However, it is instructive to consider other types of randomness.
Thus, we choose
\begin{equation}
c_k \propto e^{\imath \, \alpha_k}
\end{equation}
with constant amplitudes $|c_k|^2$ and random phases $\alpha_k$ drawn from
a uniform distribution $[0, 2 \pi]$. In Fig.\ \ref{Fig6} (a) and (b) we compare the
resulting real-time broadening of the expectation values $p_r(t)$ for this and
the previous choice of the $c_k$, where we focus on a single anisotropy $\Delta
= 1.5$ and restrict ourselves to a chain length $L=30$ to reduce computational
effort. The excellent agreement demonstrates that the specific type of randomness
does not matter. Moreover, constant amplitudes $|c_k|^2$ as such are not
responsible for the untypical dynamics observed in Fig.\ 4.

\begin{figure}[t]
\includegraphics[width=0.95\columnwidth]{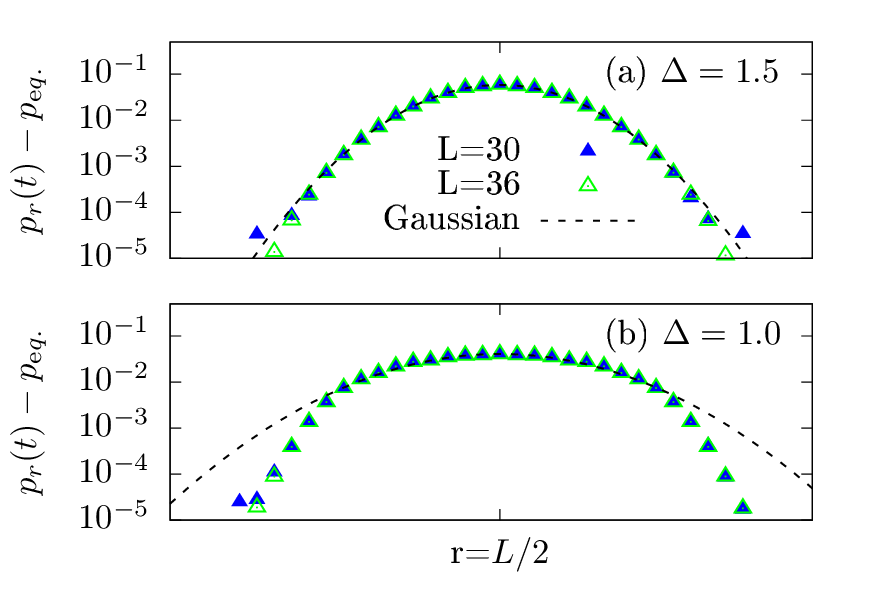}
\caption{(Color online) Density profile $p_r(t)$ with respect to site $r$ at
a single time $t J = 10$ for the two system sizes $L=30$ and $L=36$ and for the
two anisotropies (a) $\Delta = 1.5$ and (b) $\Delta = 1.0$ (symbols). Gaussian
fits are indicated for comparison (curves).}
\label{Fig7}
\end{figure}

Note that not any kind of randomness can yield the same dynamical behavior. To
illustrate this fact, let us randomize the product state in Eq.\
(8) of the main text in the following way: At all sites $r \neq
L/2$ we replace the spin-up/spin-down superposition $| \! \uparrow \rangle + |
\! \downarrow \rangle$  by
\begin{equation}
e^{\imath \, \alpha_r} \, | \! \uparrow \rangle + e^{\imath \, \beta_r} \, | \!
\downarrow \rangle
\end{equation}
with site-dependent phases $\alpha_r$, $\beta_r$ drawn from a uniform
distribution $[0, 2 \pi]$. This randomized product state has still $p_{r \neq
L/2}(0) = 1/2$ and $p_{L/2}(0) = 1$. It involves only $2 (L-1)$ random numbers,
in contrast to the state from the Haar measure with $2^L$ random numbers. In
Fig.\ \ref{Fig6} (c) we depict the resulting dynamics of the expectation 
values $p_r(t)$. Compared to the two other random cases in Figs.\ \ref{Fig6}
(a) and (b), the dynamical behavior turns out to be very different. This
difference suggests again that the lack of entanglement could be the source of
untypical dynamics.

\section{Finite-Size Effects} \label{fse}

Eventually, we show that our numerical results for the real-time broadening of
the expectation values $p_r(t)$ are free of significant finite-size effects. To
this end, we redo the $t \, J = 10$ calculations in Figs.\ 2 (a) and
3 (a) for a smaller but still large system size $L=30$. In Fig.\
\ref{Fig7} we depict the results of these calculations, together with the
previous $L=36$ data. It is clearly visible that finite-size effects are
negligibly small and are not responsible for the non-Gaussian tails at the
isotropic point $\Delta = 1.0$.


\begin{thebibliography}{}
\bibitem{note} Strictly speaking, by drawing $c_k$ at random, it is possible to get the outcome $c_k = \text{const.}$ for all $k = 1, \ldots, 2^L$.
\end{thebibliography}
\end{document}